\documentclass[11pt,twoside]{article}
\usepackage{asp2006}
\usepackage{epsf}
\def\edcomment#1{\iffalse\marginpar{\raggedright\sl#1\/}\else\relax\fi}
\reversemarginpar

\begin{document}
\title{The Pedigrees of DOGs (Dust-Obscured Galaxies)}
\author{Arjun Dey}
\affil{National Optical Astronomy Observatory, 950 N. Cherry Ave., Tucson, AZ 85719}
\author{and the NDWFS/MIPS collaboration}

\begin{abstract}
A simple mid-infrared to optical color criterion of $R-[24]\ge 14$ results in a robust selection of approximately half of the $z\sim 2$ ultraluminous infrared galaxy (ULIRG) population. These `Dust-Obscured Galaxies', or DOGs, have various properties that suggest that they are good candidates for systems in a transition phase between gas-rich mergers and QSOs.
\end{abstract}

\vspace{-0.5cm}
\section{Choosing DOGs}

\begin{figure}[!ht]
\plottwo{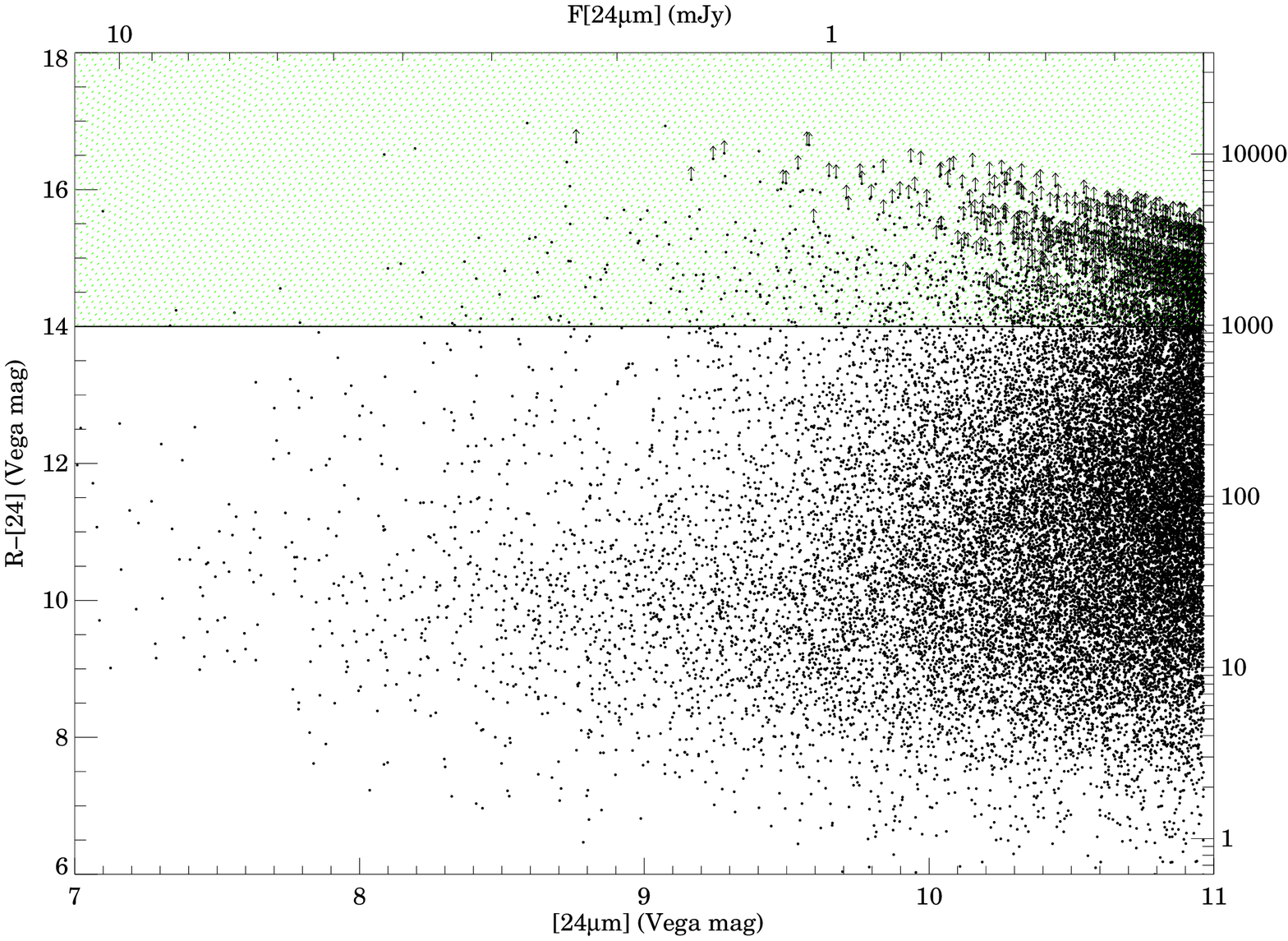}{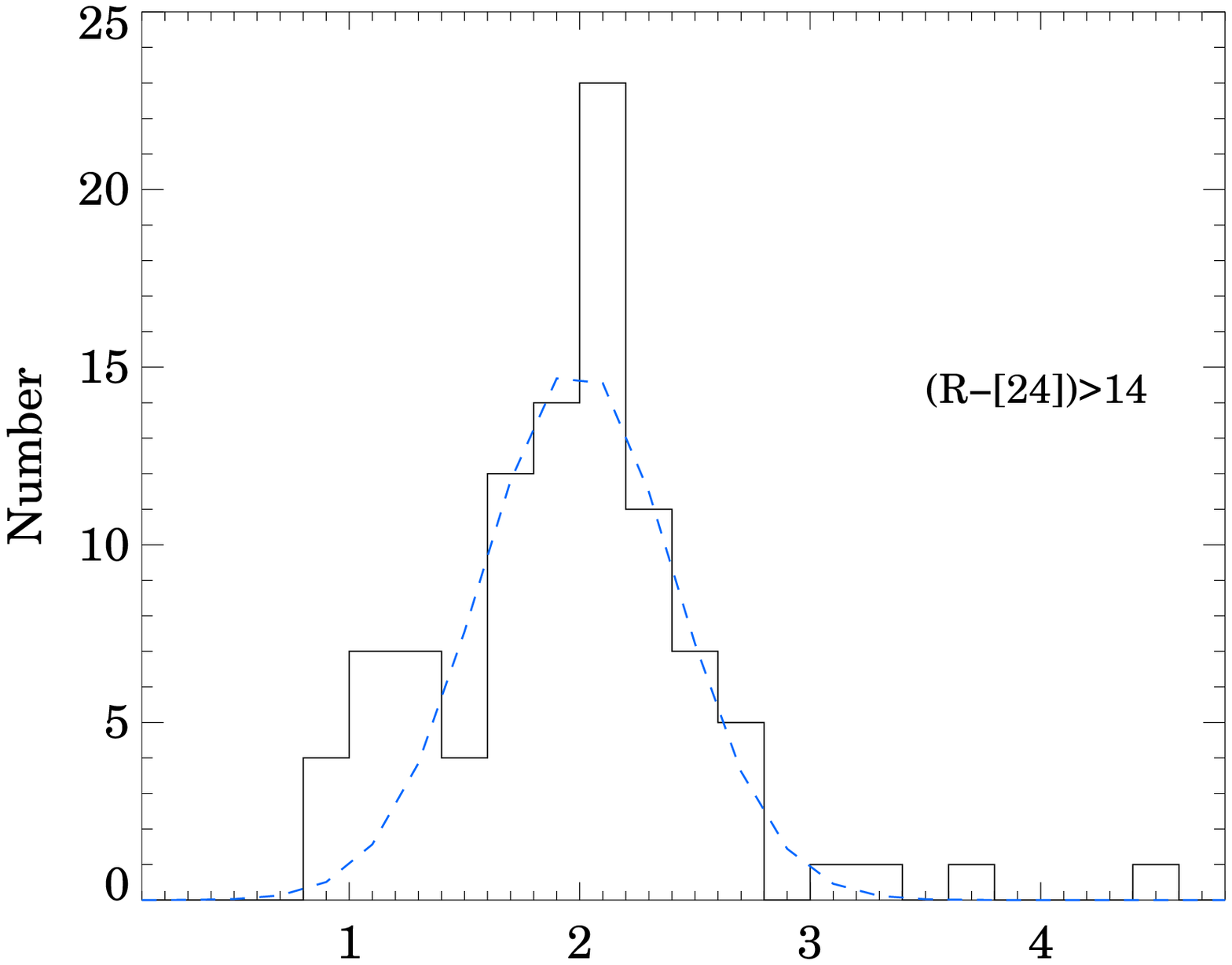}
\caption{{\it Left:} Optical/Mid-infrared color-magnitude diagram for 24$\mu$m sources in the NDWFS Bo\"otes field. {\it  Right:} Histogram of spectroscopic redshifts of DOGs in the Bo\"otes Field, obtained using spectrometers on the {\it Spitzer}, Keck and Gemini telescopes. Adapted from \citet{dey08}.
\label{colormagz}}
\end{figure}

Observations with the {\it Spitzer Space Telescope} by numerous groups have revealed a large population of ultraluminous infrared galaxies at high redshift \citep[e.g.,][and references therein]{soifer08}. In particular, our team, using the GTO MIPS 24$\mu$m {\it Spitzer} observations of the Bo\"otes field of the NOAO Deep Wide-Field Survey \citep{ndwfs} found a population of 24$\mu$m sources with very extreme optical to mid-infrared colors \citep{houck05,dey08}. Among 24$\mu$m sources with $F_{\rm 24\mu m}\ge 0.3$mJy, a small but significant fraction have colors that imply a mid-IR to optical flux density ratio $F_{\rm 24\mu m}/F_R$ in excess of 1000 (e.g., Figure~\ref{colormagz}). This fraction is only a few percent at bright 24$\mu$m flux densities, but increases to $\approx$13\% at fainter flux densities.  We refer to galaxies with these extreme colors as `Dust-Obscured Galaxies', or DOGs.  In contrast, most local ULIRGs (like Arp220) don't have such red colors, although as discussed by George Rieke at this meeting, there do exist a few local analogs (like Mrk231) that may be this red. 

Selecting galaxies based on this red color results in an excellent redshift discriminant: an $R-[24]\ge14$ (or $F_{\rm 24\mu m}/F_R \ga 1000$) selection results in 24$\mu$m sources that lie at $z\sim2$ within a fairly narrow range in redshift \citep[Figure~\ref{colormagz};][]{dey08,fiore08a}. While the color criterion is fairly arbitrary, it has the advantage of cleanly selecting $z\sim2$ ULIRGs; bluer cuts result in a broader redshift distribution that include more lower redshift, lower luminosity galaxies.  Figure 2 \citep[from][]{desai09} shows the mid-infrared colors of various samples of $z\sim2$ ULIRGs.  The simple DOG selection criterion identifies about half the ULIRGs at $z\sim2$, and the resulting population is responsible for roughly 1/4 of the infrared luminosity density at $z\sim2$.

\section{DOG SEDs}

The mid-infrared spectral energy distributions (SEDs) of DOGs span a range of spectral morphologies. At one extreme, the mid-infrared photometry is well modeled as a single power-law; at the other, the photometry exhibits a `bump', peaking roughly near 4.5$\mu$m IRAC band. The `bump' arises from the presence of the stellar continuum, peaking at (rest-frame) 1.6$\mu$m. In the sources with power-law SEDs, the emission from starlight is swamped by emission associated with the AGN. DOGs exhibit SEDs ranging continuously between these two extremes, suggesting variable contribution of the AGN relative to the starlight in the galaxies. The fraction of DOGs exhibiting power-law SEDs is a function of 24$\mu$m flux density: nearly 80\% of DOGs with $F_{\rm 24\mu m} \ga 1$mJy have power-law spectra, compared with less than 25\% of DOGs with $F_{\rm 24\mu m}\approx 0.3-0.4$mJy. 
\citet{pope08} investigated a sample of fainter DOGs in the GOODS-N field (with $F_{\rm 24\mu m}\ga 0.1$mJy) and found that their mean long wavelength SED is very similar in shape to that observed in SMGs, perhaps suggesting common origins for the cold dust emission in both populations.  Recent measurements at far-infrared and sub-mm wavelengths using MIPS aboard {\it Spitzer} and SHARC-II on the CSO have resulted in direct measurements of the bolometric luminosity of a handful of 24$\mu$-bright power-law DOGs \citep{tyler09,bussmann09b}. These measurements confirm that these power-law DOGs have $L_{\rm bol}> 10^{12}{\rm L_\odot}$ and show SEDs with significant warm dust emission, more similar to Mrk~231 than the SMGs. The present data therefore suggest that the 24$\mu$m-bright DOGs are more AGN dominated than their fainter counterparts; the latter appear to be powered primarily by star formation \citep{dey08,desai09}. 

This conclusion based on the SEDs is supported by mid-infrared spectroscopic observations with the Infrared Spectrometer (IRS) aboard {\it Spitzer}. IRS spectra of the power-law DOGs are either featureless or exhibit strong absorption by the (rest-frame) 9.7$\mu$m silicate feature \citep[e.g.,][]{houck05}. In contrast, IRS spectra of `bump' DOGs show strong PAH emission \citep[e.g.,][]{desai09,yan07}. 

\begin{figure}[!t]
\plotfiddle{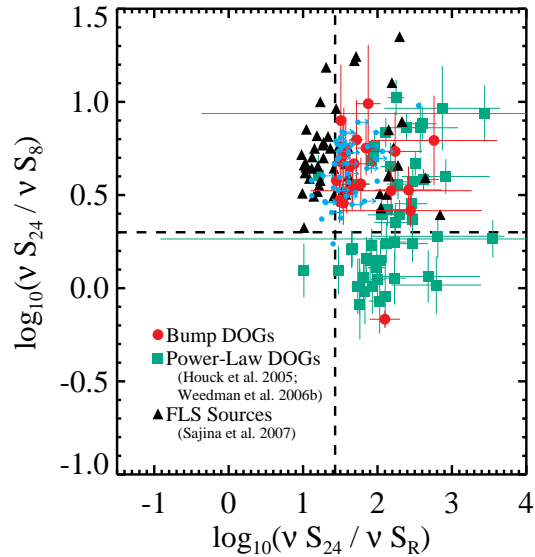}{2.6in}{0}{65}{65}{-150}{-35}
\caption{The mid-infrared colors of various samples of $z\sim2$ ULIRGs \citep[from][]{houck05,weedman06,sajina07,lonsdale09}. The vertical line represents the DOG selection criterion, which successfully identifies roughly half the $z\sim 2$ ULIRG population. Figure adapted from \citet{desai09}.
\label{iraccolor}}
\end{figure}

It is likely therefore that both AGN and star formation contribute to the bolometric luminosity in DOGs. Based on an X-ray stacking analysis of DOGs in the GOODS-S and COSMOS fields,  \citet{fiore08a,fiore08b} have suggested that DOGs may represent the long-sought population of Compton Thick AGN, and it is possible that AGN contribute some fraction of the luminosity in even the PAH-dominated `bump' DOGs \citep[although, see][for a different interpretation of the X-ray results]{pope08}.

\section{DOG Morphologies}

We have obtained high spatial resolution observations of DOGs using the {\it Hubble Space Telescope} \citep[][; also see Bussmann's contribution in these proceedings]{bussmann09a,bussmann09b} and laser guide star adaptive optics on the Keck II Telescope \citep{melbourne08,melbourne09}. These observations reveal that DOGs show a range of spatially extended morphologies. In the $V$, $I$, $H$ and $K$ bands (the rest-frame UV and optical), DOGs are spatially extended galaxies and are not dominated by a point source component.  Also, there may be a morphological distinction between the power-law dominated DOGs and the `bump' DOGs. The `bump' sources show multi-component morphologies more often, whereas the power-law sources show single peaked components with extended profiles with exponential profiles \citep{bussmann09b}. The morphologies are consistent with the power-law sources being late-stage mergers, and the `bump' DOGs being earlier stage mergers.  \citet{melbourne09} find that the more mid-infrared luminous DOGs may have more compact morphologies, consistent with the suggestion from the SEDs that the more luminous DOGs tend to be more AGN dominated, more dynamically relaxed systems.

\begin{figure}[!ht]
\plotone{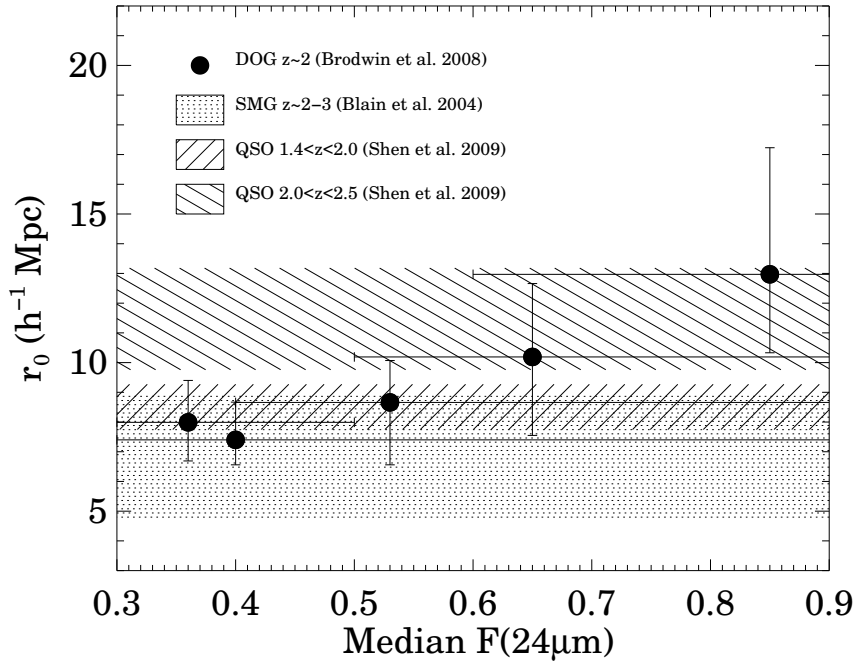}
\caption{The clustering of DOGs relative to submillimeter galaxies \citep[from][]{blain04} and QSOs \citep[from][]{shen09}. All three populations inhabit similar mass halos at the same epoch, suggesting an evolutionary link between the populations. Adapted from \citet{brodwin08}.
\label{clustering}}
\end{figure}

\section{DOG Clustering}

DOGs are strongly clustered. The clustering strengths imply halo masses of $\ga 10^{12.2^{+0.3}_{-0.2}}{\rm M_\odot}$, and support the idea that DOGs are progenitors of massive ($3-6L*$) galaxies in the present epoch \citep{brodwin08}. There is some evidence for luminosity dependent clustering among the DOG population, with the more luminous DOGs being more strongly clustered. The most luminous DOGs may inhabit halos with masses as high as $\approx 10^{13}{\rm M_\odot}$ \citep[see Figure~\ref{clustering}; adapted from][]{brodwin08}, and may therefore be the progenitors of present-day $\sim6L*$ galaxies. \citet{brodwin08} also find that the DOG halo masses are comparable to those inferred for the sub-millimeter galaxy (SMG) population. In addition, a recent measurement of the clustering of luminous QSOs by \citet[]{shen09} shows that $z\sim2$ QSOs {\it also} reside in similar mass halos (see Figure~\ref{clustering}).

\begin{figure}[!t]
\plotone{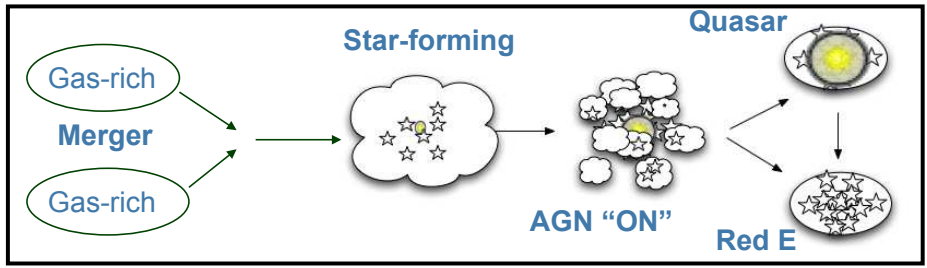}
\caption{A cartoon showing the possible role of DOGs in the evolution of massive galaxies. The initial gas rich merger results in a dust-enshrouded starburst phase (an SMG). Once the AGN is triggered, it heats the dust resulting in an increase in the mid-infrared luminosity (the DOG phase). The following evolution depends on the relative timescales of AGN fuelling, dust dissipation, and star formation, but the system could be visible briefly as a QSO before settling on the red sequence. 
\label{cartoon}}
\end{figure}

\section{DOGs and Massive Galaxy Evolution}

It is remarkable that three relatively unusual populations (DOGs, SMGs and QSOs) characterized by extreme luminosities (and therefore by short-lived phases) are found at the same epoch (i.e., $z\sim2$) and reside in similar mass halos. This strongly suggests that the three populations are different evolutionary phases in the formation of massive galaxies. 

In a seminal work, \citet{sanders88} suggested that large gas-rich mergers were responsible for the ultraluminous infrared galaxy phenomenon, and that these systems were progenitors of QSOs. In a similar vein, we speculate that SMGs, DOGs and QSOs represent phases in the evolution of a gas rich merger at high redshift (see Figure~\ref{cartoon}). The initial merger of gas rich galaxies results in a spectacular starburst, causing a rapid rise in the cold dust emission and the resulting selection and classification of the system as an SMG. A black hole is efficiently fueled in the late stages of this merger, resulting in a rapid rise in the AGN luminosity, much warmer (mid-infrared) dust emission, and the selection and classification of the system as a DOG. Finally, once the AGN has managed to clear or destroy the surrounding shrouds of dust, the source is detected as a QSO. Such a process has been described by \citet{hopkins06} as naturally explaining the origin of massive galaxies and the growth of the red sequence ellipticals. 

How could such a scenario be tested? Better gas, stellar and halo mass estimates for both DOG and SMG populations would allow an improved comparison of these two populations. If the DOG phase follows an SMG phase, then we would expect the stellar masses of DOGs to be higher than that of the preceding SMG phase, although the difference may be small (undetectable?) if the lifetimes of the two phases are brief and / or concurrent. In addition, sub-mm, mm and radio observations would provide valuable constraints on the bolometric luminosities and (through molecular line observations) gas masses. Again, if the DOG phase follows an SMG phase, then the gas-to-stellar mass ratio should be smaller for DOGs, although their dynamical masses should be comparable. We are continuing to obtain high spatial resolution imaging and spectroscopy to determine the sub-kpc morphologies and kinematics of DOGs and compare these to the SMGs.

Finally, we note that although we have measured spectroscopic redshifts for nearly 100 DOGs, an important subset lack redshifts. These are the sources for which the mid-infrared spectra are power-laws. It is possible these sources lie at even higher redshift (i.e., $z\ga 3$, where the silicate absorption feature is redshifted out of the IRS spectral window), or that the silicate feature is too weak to be detected. Deep optical and near-IR spectra are required to determine their redshifts. 

\acknowledgements
The research discussed in this contribution is based in large part on work done in collaboration with the Bo\"otes MIPS team, in particular Lee Armus, Kate Brand, Mark Brodwin, Michael Brown, Shane Bussmann, Vandana Desai, Emeric LeFloc'h, Jim Houck, Buell Jannuzi, Jason Melbourne, Alexandra Pope, Marcia Rieke, Tom Soifer and Dan Weedman. I am grateful for their insights and for allowing me to present some data prior to publication.  I am also very grateful to the organizers for inviting me to this conference and for giving me the opportunity to visit Shanghai.

\end{document}